\documentstyle[aps,12pt]{revtex}

\begin{document}
\title{Particle-hole level densities in deformed nuclei}
\author{F. Garcia$^{\ast }$, O Rodriguez \thanks{{\ }Permanent address: Instituto
Superior de Ci\^{e}ncias y Tecnologia Nucleares, Av. Salvador Allende y
Luaces, Apartado Postal 6163, La Habana, Cuba.}, F. Guzman$^{\ast }$,
H.Dias, J.D.T Arruda-Neto , and M.S. Hussein$\thanks{%
Corresponding author: hussein@axpfep1.if.usp.br}$}
\address{Instituto de Fisica da Universidade de Sao Paulo\\
C.P.63318, 05315-970 S\~{a}o Paulo, SP, Brazil}
\author{A. K. Kerman}
\address{Center for Theoretical Physics, Laboratory for Nuclear Science and\\
Department of Physics- Massachusetts Institute of Technology-Cambridge,\\
Massachusetts 02139,USA\\
\today}
\maketitle

\begin{abstract}
Microscopic Combinatorial approach is used to calculate the state and level
densities with fixed exciton numbers, in some actinide nuclei. Deformed
Saxon-Woods shell model was used as a basis from which all posible
configurations were generated. The pairing interaction was taken into
account by applying the BCS theory to each configurations. Both the spin and
parity distributions were obtained,considering the deformation of the
nucleus. Relevance of the result to parity nonconservation studies involving
epithermal neutrons on $^{238}$U and $^{232}$Th is discussed.


\vspace*{5mm}
\end{abstract}

\section{\ Introduction}

Adequate description of the nuclear level density of an excited nuclear
state with a fixed number of quasiparticles is a basic ingredient of the
statistical analysis of nuclear reactions. On the other hand, it is well
known that the exciton dependent state density used in the various
pre-equilibrium models is lacking in considering the adequate particle-hole
combination and its relation with the particular structure of the nuclei
involved \ in the reactions.

Analytical expressions for particle-hole excitation density can be obtained
by means of methods of combinatorial or statistical mechanics in terms of
the nointeracting particle model using the equidistant spacing approximation
for the single-particle states. The main deficiency of the formulae commonly
used to estimate exciton level densities comes from the assumption of
equidistant spacings for the single particle levels \cite{Gad92}\cite{Sco84}%
. Usually only few-exciton configurations contribute significantly to
pre-equilibrium emission and, in some cases , the population of low energy
configurations is relevant as well. Therefore statistical approaches and
equidistant spacings for the single particle levels do not seem to be
adequate for pre-equilibrium calculations in some instances when shell
structure effects are observed \cite{Shl95}.

In a recent work \cite{Hus97}, two of the authors studied fundamental
symmetry breaking on nuclear reaction and suggested that the observation of
sign correlations in the longitudinal asymmetry of polarized neutron
scattering from $^{232}Th$ at epithermal energies may constitute the first
direct evidence for local 2p-1h doorways. The discovery of sign correlation
in parity non-conserving (PNC) epithermal neutron-induced compound-nucleus
reactions involving the heavy nucleus $^{232}Th$ have prompted intensive
theoretical discussion concerning their origin. The great interest in the
TRIPLE\ data \cite{fran91}, stems from the fact that the statistical theory
of these reactions, although it predicts the possibility of large PNC at
individual resonances as the data also show, rules out any sign correlations
in the longitudinal asymmetry. contrary to what the data show. More recently 
\cite{Mit97}, data were obtained with smaller error bars for $^{232}Th$ and $%
^{238}U$. Feshbach and collaborators \cite{Hus97}, started with the
hypothesis that a very natural mechanism that could account for this sign
correlations is to assume that the compound nuclear PNC process occurs
through a single isolated p-wave local doorway which coupler to nearly
s-resonances. In such studies, an important quantity which could determine
how isolated the p-wave resonances are, is the 2p-1h density of states. It
has been a common practice to use the equidistant model to calculate the
np-nh level densities, which is reasonable in spherical nuclei. No thorough
investigation of the case of deformed nuclei is available. Sire nuclei such
as U and Th are strongly deformed, one would be tempted to calculate the
exciton level densities more accurately using a deformed mean field.

Few years ago M. Herman and G. Reffo \cite{Her87a},\cite{Her88} evaluated a
realistic few quasiparticle state densities with fixed exciton numbers using
combinatorial calculations in spherical and some deformed nuclei. Their
calculations were carried out in the basis of single particle orbitals
derived from the harmonic oscillators defined by parameter due to Seeger and
Howard \cite{See75}. Calculation of spin-dependence particle-hole level
density were carried out only for spherical nuclei. For these nuclei the
Gaussian-Wigner type spin factor was used. A direct test of the validity of
this procedure is afforded by calculations made by Herman and Reffo. They
found that the Gaussian formula works very well for configurations
containing at least three excitons. However, in the case of deformed nuclei
there are no comprehensive calculations with take into account the role of
the collective rotational degrees of freedom for the evaluation of the
particle-hole level density, with the question of the spin factor still open.

It is generally assumed, that levels of different parities have an
equiprobable condition. This is the Ericson's prediction \cite{Eri60} for a
non-interacting many fermion system. The possibility of considerable
irregularities occurring in the distribution of positive-and negative-parity
levels is demostrated in particular by the results of combinatorial
calculations of level density carried out within the framework of the
non-interacting particle model \cite{Hil69} as well as those based on the
more rigorous quasi-particle-phonon model, which takes into account the
collective excitations of nuclei \cite{Sol74} \cite{Vdo76}. Previous
combinatorial microscopical calculations \cite{Her87a} show that the results
have evidently oscillatory character around the equal probability value. The
fluctuaction are observed throughout a wide energy range ( \symbol{126}30
MeV),even though on the average the amplitude of the fluctuactions decreases
with increasing excitation energy. The fluctuactions are reduced with the
increase of the deformation, exciton number and mass of the nucleus. It is
an indication that in most of the applications, the assumptions of equal
parity distribution seems, to be justified. However, certain channels,
particularly sensitive to the parity distribution (e.g. radiative channels),
have to be treated with caution.

Another interesting question is related to the deformation parameters used
in the calculation of single-particle spectra. Herman et al in \cite{Her88}
studied the effect of nuclear deformation on few-quasiparticle
state-densities. It is shown that nuclear deformation tends to suppress
strong fluctuations observed in the similar calculations performed in the
space of spherical shell model basis vectors. The conclusion of Hermann and
Reffo have the cualitative character, since some fine details of particular
structure of studied nuclei ($^{27}$Al, $^{100}$Mo, $^{170}$Er, ) are not
taken in to account. \ In our view, the first step should be the right
selection of deformation parameters with the calculations of single particle
spectrum in the equilibrium deformation. The study of fine nuclear structure
effects, of order of PNC\ need the most accurate calculation of particle
hole level density, which is very sensitive to the single-particle spectrum.

The paper is devoted to the development of a more accurate method for the
calculation of particle-hole level densities for deformed nuclei by
thoroughly considering the rotational degree of freedom. Particular effort
is made in applying the method to actinide nuclei used in parity
no-conservation studies induced by epithermal neutrons. The final intent is
to evaluate the various uncertainties, constraints or difficulties related
to the level density question and their impact on nuclear reaction
predictions.

In sect.2 we outline very briefly the formalism with the description of the
general method which improves our previous mode. The preliminary treatments.
The problems of the spin and parity distributions is presented in sect.3. In
sect.4 we discuss some results for two nuclei belonging to the actinide
region. Finally, we present the main conclusions in sect.5.

\section{\protect\bigskip Particle-hole level density calculation's Method}

In order to calculate the 2p-1h doorway states densities coupled to total
angular momentum J we used the microscopic combinatorial method \cite{Her87a}
\cite{Her88}\cite{Gar93} at excitation energy E, considering the axial
deformation of the compound nucleus. The calculation was performed in the
space of realistic shell-model single particle states, including pairing
interaction in terms of the BCS theory.

The first step to start the single-particle spectrum calculation is to fix
the value of the equilibrium deformation of the nucleus using an optimal set
of potential parameters\cite{Che68}. In the shell model approach, based on
mean-field potentials, the Strutinsky method is usually used\cite{Str67}. We
have calculated the extreme points of the potential energy along the path
from the equilibrium deformation to the more unstable form, using the
BARRIER code \cite{Bar98} which includes the Strutinsky prescription and the
Pashkevich parametrization for the nuclear shape\cite{Pas71}. The
calculation was performed taking into account quadrupole and hexadecapole
deformation parameters, which is a reasonable approximation to describe the
equilibrium ground state. It is important to note that for more complex
nuclear shapes, higher order deformation parameters should be included in
the calculations\cite{Gar98a}.

The other quantity which needs to be accurate by determined is the pairing
strength parameter G$_{n,p}$. There are few methods commonly used to fix the
value of G. Some authors \cite{Mai80}, \cite{Ben81}, take the pairing
strength as a free parameter chosen to fit the calculated level densities to
the experimental number of resonances at the neutron binding energy. This
procedure certainly gives the appropriate level densities, but the pairing
strength parameter derived this way is not related directly to the pairing
interaction and may hide several inadequacies of the applied theory. We have
chosen the method of relating the G parameter to the difference of masses of
neighboring nuclei.

In order to obtain intrinsic state densities all configurations with a
specified number of particles and holes are generated within the assumed set
of the single-particle levels by means of the permutation enumeration
algorithm. For each of them a proper coupling of the spin projections is
performed to obtain the nuclear states. The state density w(E,M) is found by
counting states with the angular momentum projection M falling in the 0.5
MeV interval centered at the excitation energy. The pairing interaction was
considered by applying the BCS model to each configuration so that a better
understanding of the pairing correlations in few quasiparticle states could
be achieved.

We also used an improved formulae for the spin factor of particle-hole level
density. Owing to the axial-symmetry of the ground states of these deformed
nuclei we will use the spin factor formula which contains the rotational
degree of freedom contribution, proposed by S. Bjornholm, A.Bohr and
B.R.Mottelson in Ref\cite{Bjo73}. The statistical microscopical method \cite
{Gar93} is used for to obtaining the parallel and perpendicular spin-cutoff
factors in a consistent way.

\subsection{The single-particle spectra}

In the last few years, several calculations of single particle level schemes
have been carried out both as a function of nucleon numbers and as a
function of nuclear shape in connection with the studies of nuclear
deformation potential energy surfaces. Simple global parameters for the
underlying model potential have also been estimated on the basis of fits to
nuclear ground state masses and fission barrier heights. Based on these and
other single particle level schemes and the partition function approach,
several numerical calculations of nuclear level densities have been carried
out.

In this work the nuclear shape parametrization is carried out using the
Cassini parametrization \cite{Pas71} According to this the deforming shape
(up to and beyond its separation into two fragments) can be conveniently
described by the Cassini ovaloids figures

Considering only axially symmetric nuclear shapes, the Cassini ovaloids are
taken as the first approximation to the nuclear shape. The deviation from
the ovaloid shape is given by an expansion into a series of Legendre
polynomials. Geometrically, the family of Cassini ovaloids is defined by 
\cite{Has88}:

\begin{equation}
r^{2}(z,\epsilon )=\sqrt{(a^{4}+4(cz)^{2})}-(c^{2}+z^{2}-\epsilon ^{2}).
\label{radius2}
\end{equation}
In this equation, r and z are cylindrical coordinates; $\epsilon $ is a
dimensionless quantity such that c=$\epsilon R_{0}^{2}$; {\em c} stands for
the square distance from the focus of the Cassini ovaloids to the origin of
coordinates; and $a$ is a dimensionless parameter which completely defines
the shape taking into account volume conservation.

In the plane containing the symmetry axis one can define a system of
coordinates $(R,x)$ such that the coordinate line $R$ is constant. This is a
Cassini ovaloid, where $0\le R<\infty $ and $-1\le x\le 1$. The $(R,x)$
coordinates are related to the cylindrical ones $(r,z)$ by the following
equations

\begin{equation}
R(z,r)=\sqrt[1/4]{[(z^{2}+r^{2})^{2}-2\epsilon R_{0}^{2}\cdot
(z^{2}-r^{2})+\epsilon ^{2}R_{0}^{2}]}  \label{RadiusZr}
\end{equation}

\begin{equation}
x(z,r)=\frac{sign(z)}{\sqrt{2}}\left[ 1+\frac{z^{2}-r^{2}-\epsilon R_{0}^{2}%
}{R^{2}(z,r)}\right] ^{\frac{1}{2}}  \label{x}
\end{equation}

In this system of coordinates, the basic shape of the nucleus is described
by these equations, where $R$ is constant, determining the Cassini ovaloids.
Thus, the nuclear shape can be defined as a curve $R(x)$ that does not
intersect any straight line $x=constant$ at more than one point.
Accordingly, we can expand the function $R(x)$ into multipoles, giving

\begin{equation}
R(x)=R_{0}[1+\beta _{m}Y_{m0}(x)].  \label{Rx}
\end{equation}

The set of parameters ($\epsilon $, $\beta $) completely determine the
nuclear shape. The details of this parametrization are given in \cite{Bar98}%
. As an example, we show in Figure 1 $\{\epsilon ,\alpha _{4}\}$ as
functions of $\{\beta _{2},\beta _{4}\}$. As is clearly seen in this figure,
it is difficult to establish an analytical connection between the two set of
parameters. This relation was obtained by a least--square fit of the
parameters $\beta _{2}$, $\beta _{4}$, for the harmonic spherical expansion,
to our shapes described by the Cassini ovaloids. Using this figure it is
possible to establish a connection between the two set of parameters to
describe the same nuclear shape, but for more complex shapes more
coefficients are needed in the harmonic spherical expansion.

In order to obtain the single--particle energies and wave functions, the
Hamiltonian matrix elements are calculated with the wave functions of a
deformed axially symmetric oscillator potential. The basis cut-off energy is
determined in such a way that the negative energy eigenvalues of the
Woods--Saxon potential do not change when adding more harmonic oscillator
shells.

As usual, the real potential $V(r)$ is expected to follow roughly the
density distribution. One of the most used radial dependences, is that of
the Woods--Saxon potential, which takes into account the nuclear potential
and density distribution. This potential involves the parameters V$_{0}$, $%
r_{0}$ and $a$, describing the depth of the central potential, the radius
and the diffuseness parameters, respectively.

A definition of the radial dependence of the potential for a deformed
nucleus, with an arbitrary shape of the surface, was proposed by Pashkevich 
\cite{Pas71}. According to Pashkevich, the nuclear potential is given by

\begin{equation}
V(r,z,\epsilon ,\hat{\beta})=\frac{V_{o}}{1+exp^{\frac{dist(r,z,\epsilon ,%
\hat{\beta )}}{a}}}  \label{vrz}
\end{equation}
$\label{vr}$where dist(r,z,$\epsilon ,\hat{\beta}$) is the distance between
a point and the nuclear surface, $a$ is the diffuseness parameter and V$_{0}$
the depth of the potential well.

The Woods-Saxon-type potentials, with the spin-orbit interaction
proportional to the potential gradient, are the most appropriate from the
physical point of view. The spin-orbit interaction is given by:

\begin{equation}
V_{so}(r,z,\epsilon ,\hat{\beta})=\lambda \left( \frac{h}{2Mc}\right)
^{2}\nabla V(r,z,\epsilon ,\hat{\beta})\cdot (\vec{\sigma}\times \vec{p})
\label{vso}
\end{equation}
where $\lambda $ denotes the strength of the spin--orbit potential and M is
the nucleon mass. The vector-operator $\vec{\sigma}$ stands for Pauli
matrices and $\vec{p}$ is the linear momentum operator.

The Coulomb potential is assumed to be that corresponding to the nuclear
charge $(Z-1)e$, and uniformly distributed inside the nucleus.. In short,
the depth of the central potential is parametrized as

\begin{equation}
V=V_{0}[1+\kappa ^{\prime }T_{z}],\text{ \ }\kappa ^{\prime }=\frac{2k}{A}
\label{v}
\end{equation}
with T$_{z}$ is the z-component of the isospin and k is an adjustable
parameter.

The single-particle spectra were obtained by means of the new\ version of
the BARRIER code \cite{Bar98}, based in a WSBETA code \cite{Cwi87}, using
axial deformed Saxon-Woods nuclear potential well defined by parameters, due
to Chepurnov \cite{Che68}.

\subsection{Microscopic Combinatorial approach for level densities with
fixed exciton numbers}

The combinatorial method to obtain the level densities provides the
possibility of direct counting of the levels with a fixed number of particle
and holes. We start from the finite set of single particle states derived
from the shell model with appropriately nuclear model potential, whose
detailed description has been made above. Usually 100 neutron and 100 proton
orbitals were used in our calculation. It allows us to calculate the
particle-hole state densities theory for deformed nuclei as Th$^{233}$ and U$%
^{238}$, in energy range below 10 MeV.

The pairing interaction was taken into account by applying the BCS theory to
each configurations. All residual interactions but pairing are neglected.

The total energy of each configuration is determined through the
superconductivity theory. Configuration dependence is introduced into BCS
theory by the blocking method as proposed by Wahlborn \cite{Wah62}, to allow
for blocking of more than are orbital.

Accordingly for each generated configuration, a set of two BCS equations is
solved 
\begin{equation}
\begin{array}{c}
N=2\sum_{i}^{\prime }\,\,U_{is}^{2} \\ 
{\frac{2}{G}\,\,=\,\,\sum_{i}^{\prime }\,\,[(\varepsilon _{i}-\lambda
_{s})^{2}+\Delta _{s}^{2}]}^{-1/2}
\end{array}
\label{eqbcs1}
\end{equation}
\noindent where 
\begin{equation}
U_{is}^{2}\,\,=\,\,\frac{1}{2}\,\,\left( 1-\frac{(\varepsilon _{i}-\lambda
_{s})}{[{(\varepsilon _{i}-\lambda _{s})}^{2}+\Delta _{s}^{2}]}\right)
\,\,\,\cdot  \label{eqbcs2}
\end{equation}
\noindent Here $\varepsilon _{i}$ is the single particle energies, N stands
for the number of paired nucleons, and $\lambda _{s}$ and $\Delta _{s}$ are
the chemical potential and correlation function for a given configuration
that are supposed to be determined.

The total configuration energy, according to the BCS model 
\begin{equation}
E_{s}\,\,=\,\,\sum_{J}\,\,\varepsilon _{i}\,\,+\,\,2\sum_{i}^{\prime
}\,\,U_{is}^{2}\,\,\varepsilon _{i}\,\,-\,\,\frac{\Delta _{s}^{2}}{G}
\label{eqbcs3}
\end{equation}

The excitation energy is calculated in turn as the difference between the
total energy of a configuration and the total energy of the ground state
where the first summation included only blocked orbitals.

The particle-hole level density were obtained by means of the ICAR\ and CONV
codes \cite{Her87b}

\subsection{Collective degrees of freedom. Spin dependence}

Collective phenomena in nuclei are receiving considerable attention in the
analysis of spectroscopic data on the characteristics of low-lying levels.
Various microscopic methods of describing the structure of collective levels 
\cite{Mig65}\cite{Sol71}\cite{Sol89}are also widely used at present to
consider the interrelationship between the collective excitations and the
single-particle motion of nucleons in a self consistent nuclear potential.

Strictly speaking, any separation of collective variables should be
accompanied by a corresponding decrease in the number of internal degrees of
freedom. But since collective motions are formed owing to deep-lying
nucleons, while internal excitations are determined basically by the
single-particle levels adjacent to the Fermi surface, exclusion of the extra
degrees of freedom in the low-temperature region should not strongly affect
the density of the internal excited states. Under these conditions,
adiabatic consideration can be fully justified, at least as a first step in
the analysis of the rotational increases,in the level density of nuclei.

The impact of different models for spherical nuclei on the consideration of
collective enhancement has been studied in Refs.\cite{Bor-lib}\cite{Eis87}.
Herman and Reffo \cite{Her87a},have considered only internal degree of
freedom at low energies without account of the rotation and the vibration
enhancement can not be considered for these heavy nuclei The validity of the
statistical law describing the spin distribution of nuclear levels must be
reconsidered when applied to levels with fixed exciton numbers.

For spherical nuclei the formula for the spin distribution function reads 
\begin{equation}
R(J)\,\,=\,\,\frac{2J+1}{2(2\pi )^{\frac{1}{2}}\sigma ^{3}}e^{-[{(J+\frac{1}{%
2})}^{2}/2\sigma ^{2}]}  \label{SpheRj}
\end{equation}
which is derived under the assumption of a Gaussian distribution of spin
projections M. While it is very likely to be true when the number of levels
is high enough, this assumption may not hold for levels with low exciton
numbers, for which the density of states is too low for statistical
treatment, to be justified.

In the case of spherical nuclei Herman and Reffo \cite{Her87a} found that
Eq. (5) is valid for levels containing at least four and, to some extent,
also three excitons.

The collective contribution to the level density of a deformed nucleus is
defined by the symmetry order of nuclear deformation. For deformed nuclei
the spectrum of its energy states will be determined not only by the
internal excitations but also by the rotation of the nucleus as a whole.
This rotation may lead to a considerable increase in the density of nuclear
levels. Since the deformation of the nuclear potential removes the
degeneracy of the basis vectors belonging to the same spin multiplet, we are
no longer able to obtain the spin distribution of nuclear levels, according
to Eq. (5).

Reffo et al \cite{Her88} have studied the influence of nuclear deformation
on the distributions of quasiparticle states. It is shown that deformation
tends to suppress the strong fluctuations observed in the similar
calculations performed in the space of spherical shell-model basis vectors.
In a deformed nucleus, each intrinsic state gives rise to a rotational band
on the total level spectrum, for a given angular momentum. It is therefore
obtained by summing over a set of decomposition of the level spectrum, as
for a spherical system. The experimental level data of the nuclei under
study show that this prediction is well established. In that case, the
following expression has been used for level-density calculation of
axially-symmetric nuclei 
\begin{equation}
\begin{array}{c}
R(J)\,\,=\,\,\frac{2I+1}{\sqrt{8\pi }\,\sigma _{\parallel }}\,\,\,e^{-\left( 
\frac{-I(I+1)}{2\sigma _{\perp }^{2}}\right) } \\ 
\rho _{2p1h}\,\,(U,J)\,\,=\,\,\omega _{2p1h}(U)\,\,.\,\,R(J)\,\,\,\cdot
\end{array}
\end{equation}
\noindent In this formula the contribution of rotational states is taken
into account. Here $\omega _{2p1h}$ (U) is the microscopic level density of
2p-1h states.

The spin cut off parameters $\sigma _{\perp }^{2}$ and $\sigma _{\parallel
}^{2}$ are calculated in the following way \cite{Den93}. 
\begin{equation}
\sigma _{\perp }^{2}=\frac{\Im _{\perp }t}{\hbar ^{2}}\,\,\,\,\,\,\,\,\,\,\,%
\,\,\,\,\sigma _{\parallel }^{2}=\Omega ^{-2}\,gt\,\,\,\cdot
\end{equation}
\noindent where $t$ is the nuclear temperature and g is the single-particle
level density near the Fermi energy, and $\Omega ^{-2}$ is the value of the
average single-particle square projection on the symmetry axis of deformed
nucleus.

The energy dependence of the moment of inertia $\Im _{\perp }$ is
approximated in the following way: 
\begin{equation}
\Im _{\perp }\,=\,\{ 
\begin{array}{cc}
(\Im _{0}-\Im _{rig})\left[ 1-\frac{U}{U_{crit}}\right] & U<U_{crit} \\ 
\Im _{rig} & U>U_{crit}
\end{array}
\end{equation}
\noindent $\Im _{0}$ is the moment of inertia in the ground state, U$_{crit}$
is the maximum value of the transition energy from the superfluid to the
normal state for neutron and proton system, and $\Im _{rig}$ is the rigid
body moment of inertia of the nucleus ($\Im _{rig}=\frac{2}{5}Mr_{o}^{2}A^{%
\frac{2}{3}})$.

\subsection{The parity distribution}

According to Ericson\cite{Eri60}, even and odd parity levels contribute
equally to the level density. Being aware of the fact that the equal parity
distribution is very questionable for levels with a fixed number of exciton,
we have performed combinatorial calculations to investigate this problem.
This distribution at a fixed value of energy can be described in terms of
the asymmetry ratio:

\begin{equation}
A(U)=\frac{N^{+}(U)}{N^{+}(U)+N^{-}(U)}
\end{equation}

where N$^{\pi }$ are the number of levels with parity $\pi =\pm 1.$

\section{\protect\bigskip Calculations and Results}

Based on the above described procedure, we consider in the following $^{238}$%
U and $^{232}$Th. Both nuclei are well deformed in their ground states and
therefore, when they interact with epithermal neutrons the compound nucleus
acquire reasonably high deformations. For this reason, these two nuclei
constitute a natural test of the procedure proposed in the present work.

Figures 1a and 1b show the potential energy surface calculated in the frame
of the Strutinsky's Method \cite{Str67} and the corresponding ground state
deformation points for each case which are in agreement with the
experimental data. As already mentioned, a comprehensive description of the
level densities should include the characteristic of the nuclear
configuration for different deformation parameters. In this sense, we show
also other extreme points which define a possible path in the
multidimensional space of the deformation degrees of freedom. This path
could describe a decay-process of the compound system from its ground state
and therefore has to be considered in calculations of cross sections in
particular in reactions induced by epithermal neutrons.

Usually in calculations dealing with such highly deformed and excited
compound nuclei where a drastic change of the nuclear shape ensues, one
relies on extrapolations from the ground state structure properties. Of
course, this represents a very crude procedure because in these regions the
shell closure conditions are changing and one has to take into account the
role of \ higher multipole coefficients to describe the nuclear shape.

The Chepurnov parameters, and the obtained deformation parameters allow us
to reproduce the experimental low-lying quasi particle states of the nuclei
and the ground state of neighboring nuclei.

The extremal points in Potential Energy Surface of Th$^{233}$ are given in
Table 1.

\ The single particle level density is a basic ingredient in the
calculations, because the single particle spectra depend on the symmetry
properties of the potential well. In nuclei deformed in their ground states
the influence of the second minimum on the level density will be sharply
decreased because of \ a smaller difference between deformation in two
minima. This situation is showed in Table 2.\ 

The calculations of the particle-hole level densities were carried out using
the modified version of ICAR code \cite{Her87b}. The results of the
calculation for 2p-1h states are shown in Fig.2 for positive parity in
comparison with the Williams formula. The solid line represents the Williams
formula, while the dashed lines corresponds to calculation obtained with our
method using the Cherpunov parameters of Wood-Saxon nuclear potential. The
agreement is reasonable, though the Williams formula represents an average
of the combinatorial level density.

The pairing strength parameter G is calculated from the difference of masses
of neighboring nuclei as proposed by M. Herman \cite{Her87a}. The values
obtained for $^{232}$Th and $^{238}$U are given in Table 3.

In Fig.3 we show the comparison of the behavior of the spin cutoff
parameters $\sigma _{\perp }^{2}$ and $\sigma _{\parallel }^{2}$ as function
of energy, for Th$^{233}$ and U$^{239}$. The solid lines represent the
results for the Th isotope using the approach described in eq.(13), while
the dashed lines correspond to the calculations for the U isotope. In the
case of $\sigma _{\perp }$ the figure shows a similar behavior for both
isotopes, which reflects the similarity of the moments of inertia calculated
in our approach; this results from almost equal shapes found in the ground
state. In the case of $\sigma _{\parallel },$ the calculations exhibit \
marked differences for the two isotopes. This fact reveals the need of
taking into account the single-particle spectrum for each particular
isotope, since for $\sigma _{\parallel }^{2}$\ this plays a crucial role.

In Fig.4 we show the behaviour of the spin factor distribution as a function
of energy, for Th$^{233\text{ }}$in the spherical and the deformed
approaches. Note that the statistical assumption of a Gaussian distribution
with spherical symmetry is not always justified, because for the cases of
low J (in Fig.4 we show for J=$\frac{1}{2}$, see also Figs.5 \ and 6) it
underestimates the calculations, which has an important influence on
analyses such as those performed, in \cite{Hus97}. In Fig.5 we compare the
influence of the spin cutoff parameters on the surface of the distribution
R(E,J), in comparison with the spherical formulae using eq.(11) for low
energy and J-values. The absolute values differ by nearly two orders of
magnitude for the spherical and deformed cases.

For small p-h excitations the exact combinatorial solutions described in
Sect.II should be used. To take into account the rotational degrees of
freedom, the rotational bands are built on each quasiparticle state, using
the average values of the rotational constants A$_{r}$ and B$_{r}:$

\[
E(J,K^{\pi })=E(K^{\pi })+\left[ J\left( J+1\right) -K^{2}\right]
A_{r}+B_{r}\left( J+\frac{1}{2}\right) \left( -1\right) ^{J+\frac{1}{2}%
}\delta _{K,\frac{1}{2}} 
\]

The low energy region is interesting from the point of view of its
difference from the region where statistics is applicable. The number of
levels is small and their energy and quantum number distributions may have
strong fluctuations.

For large p-h excitations, the p-h level densities with spin distribution
should be used with the approach described in Eq.(12). The width of the spin
distribution can be determined using Eqs.(13) and (14). This procedure
provides a convenient way for the analysis of deformed nuclei.

The results of the microscopic calculations of two-particle-one hole level
densities for J=$\frac{1}{2}$, in Th$^{233}$ and U$^{238}$, are shown in
Fig. 6. The impact of collective rotational enhancement in the level density
can result values 3 times large than those from the Williams formula using
the standard parameters for these nuclei. This procedure can be useful in
order to reevaluate the parameters of the Williams formula.

Finally, in Fig.7 the trend of the fraction of positive parity levels for
the analyzed nuclei as a function of excitation energy (up to 15 MeV) is
plotted. Here, this fraction is given by the parity asymmetry defined by
Eq.(15) as a function of U, where N$^{\pm }(U)$ is calculated by a direct
counting of the levels. The regions of lower energies show the greatest
fluctuations in A (eq. 15); although these fluctuations persist at higher
energies, a slow smooth out is observed. This observed strong oscillatory
character, not predicted by the Ericson hypothesis, must be taken into
account in studies of PNC, for which the fine structure effects are
important.

\section{Conclusion and Final Remarks}

We have presented a realistic few quasiparticle level density approach for
deformed nuclei, which should be preferred to the more phenomenological
expression commonly used. The more realistic inclusion of the rotational
degree of freedom improves the calculations in the region of the actinide
nuclei, where deformation is not negligible even in the ground state. The
analysis of nuclear reactions in the region of the actinide nuclei, with
epithermal neutrons, requires the consideration of nuclear deformations
where the amplification of the collective movement must be taken into
account. The deformation introduces significant fluctuations in the level
densities. This prevents the use of a single analytical expression able to
reproduce these densities in different mass regions. In this regard, a
judicious analysis of the extreme points becomes essential. When one needs
to analyze the shape of the nucleus corresponding to the external saddle
point, it is necessary to use a procedure like the one here proposed,
because the nuclear deformation for a fixed configuration of particles and
holes neither suppresses the fluctuations that takes place in the level
densities, nor the parity distribution. The importance of our calculation
for an eventual quantitative understanding of the mechanism of PNC in
epithermal neutron induced compound reactions using actinide nuclei is
discussed. 

\section{Acknowledgments}

This work was partially supported by the CNPq-CLAF and FAPESP, and in part
by funds provided by the U.S. Department of energy (D.O.E.) under
Cooperative Agreement \# DE-FC02-94ER40818.

\newpage

\section{Table Captions}

Table 1. Deformation parameters of extremal points in potential energy
surface of $^{233}$Th.

Table 2. Single-particle level densities of extremal points in $^{233}$Th.

Table 3. Pairing constants of $^{233}$Th and $^{238}$U.

\newpage

\section{Figure Captions}

Figure 1a. $^{233}$Th Surface Potential Energy. See text for details.

Figure 1b. $^{239}$U Surface Potential Energy. See text for details.

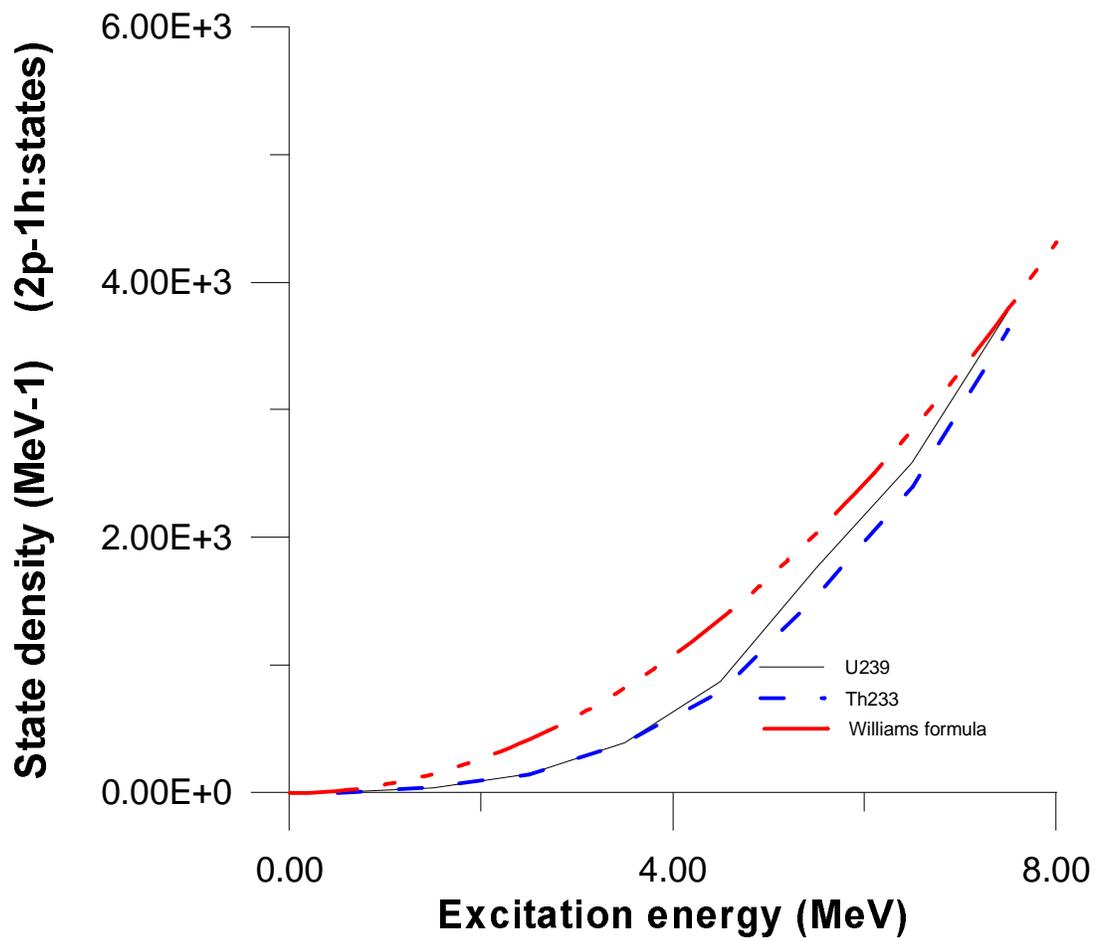
Figure 2. Microscopic combinatorial 2p-1h level density corresponding to
positive parity in comparison with the William formula. See text for details.

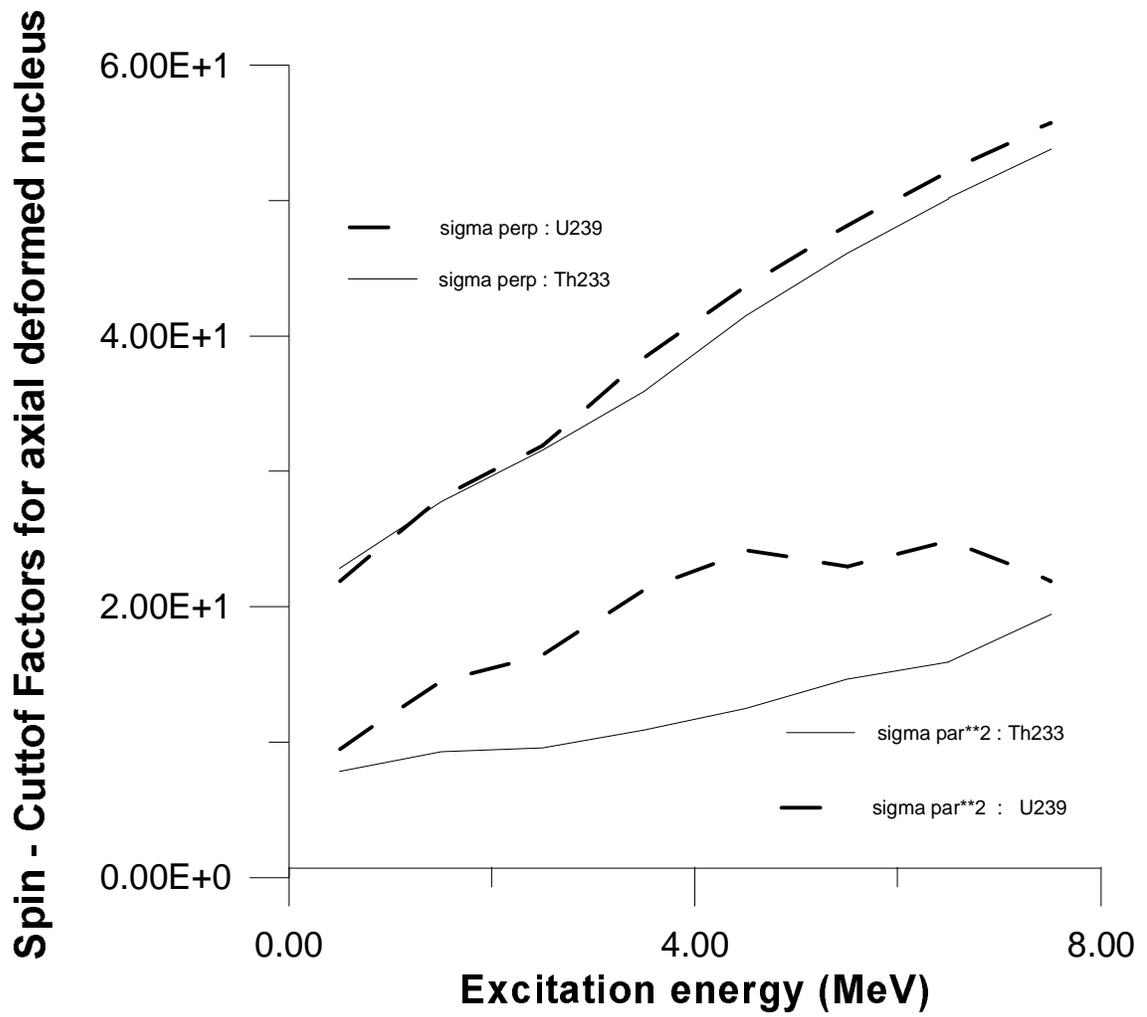
Figure 3. Spin cuttof parameters vs. excitation energy. See text for details.

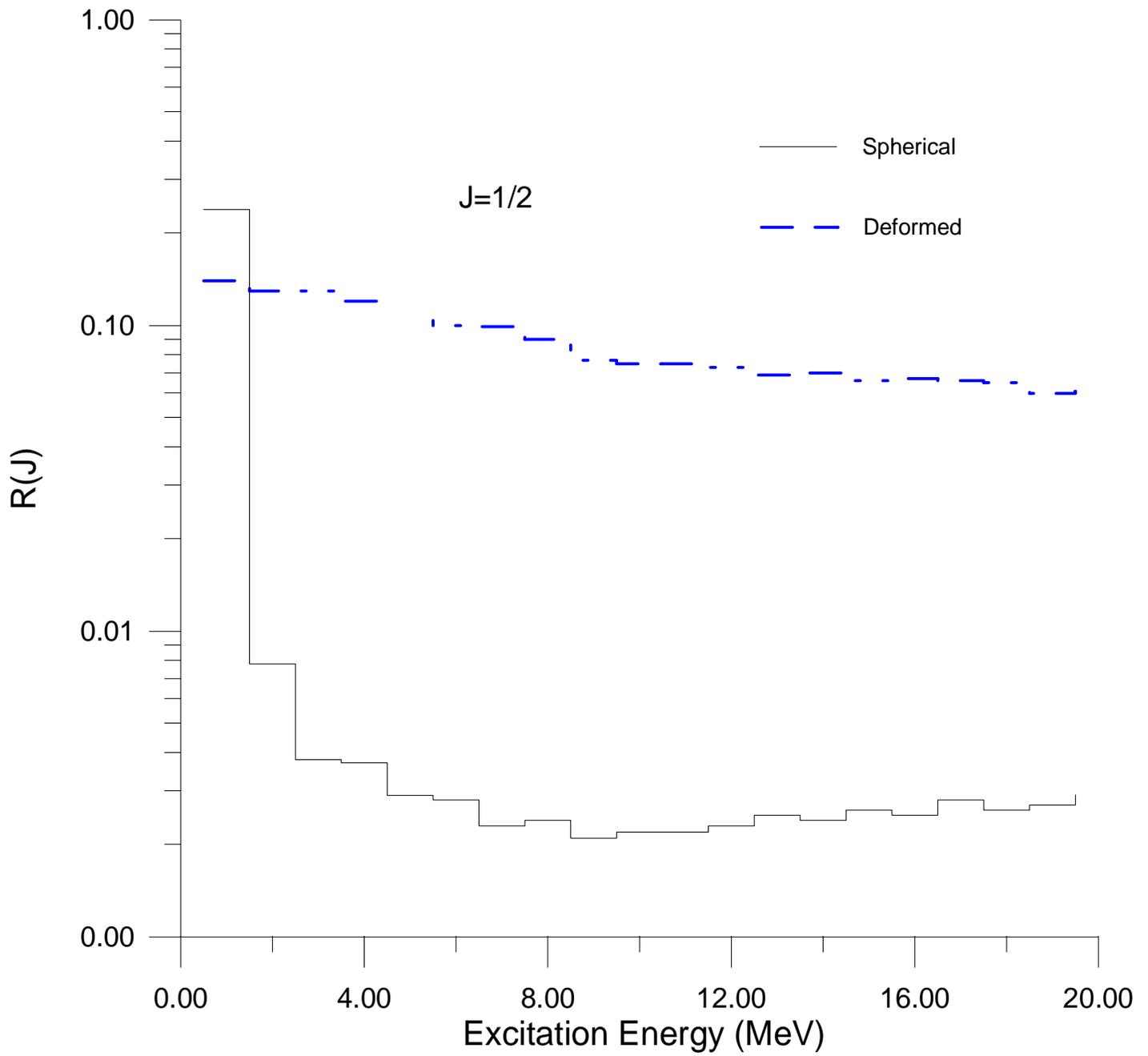
Figure 4. The level density spin distribution factor R(J) for $^{233}$Th vs.
excitation energy

Figure 5. Two-dimensional Plot of R(J) for $^{233}$Th. See text for details.

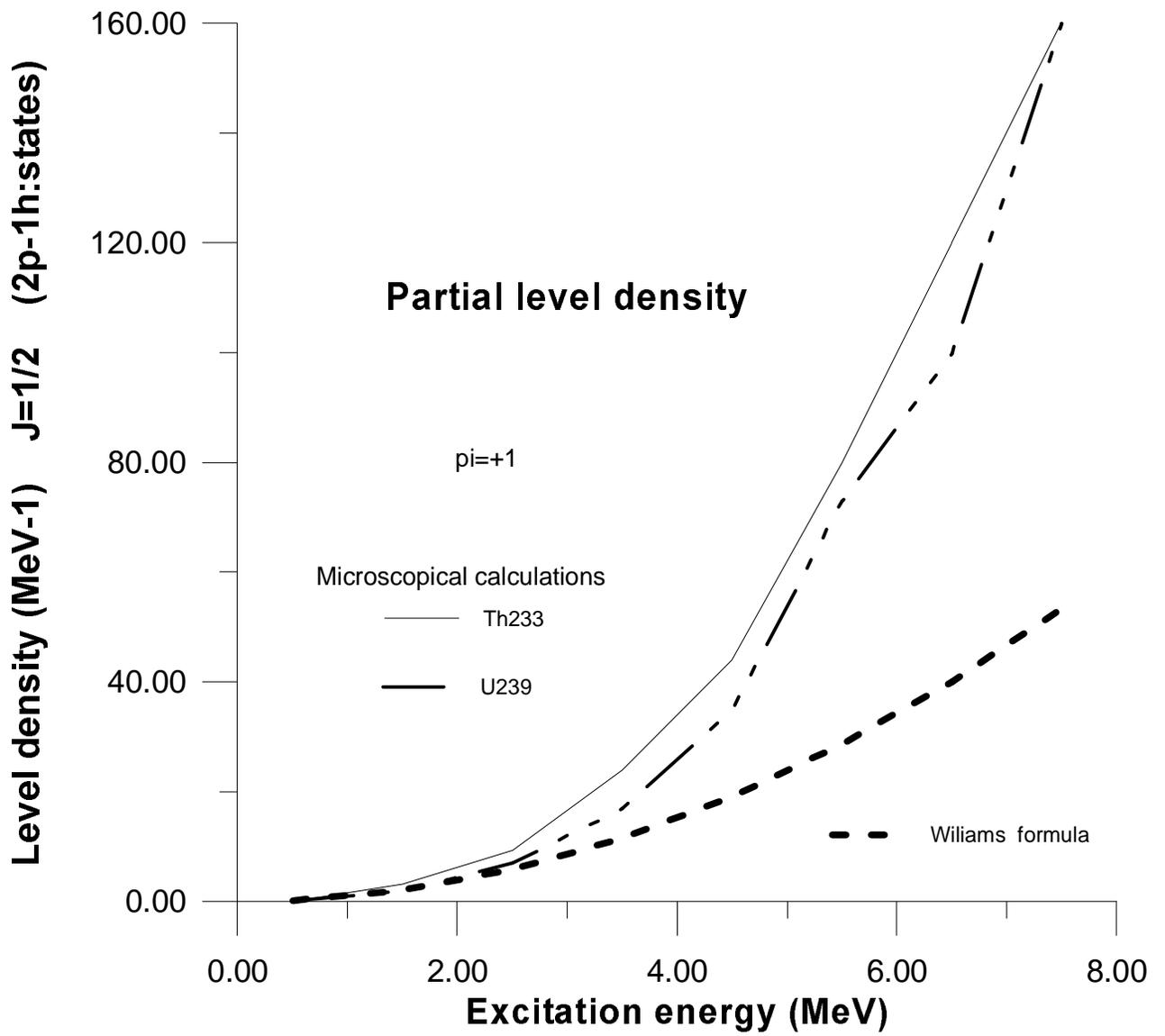
Figure 6. The 2p-1h level density for $^{233}$Th and $^{239}$U. The
Microscopic Combinatorial Method results are compared with the equidistant
Williams formula.

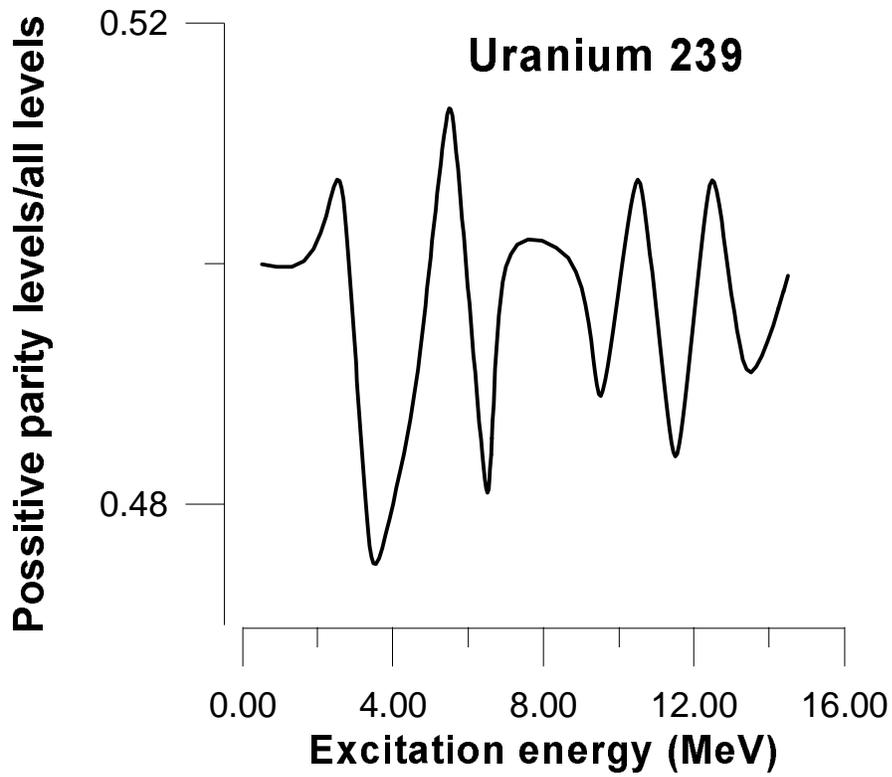
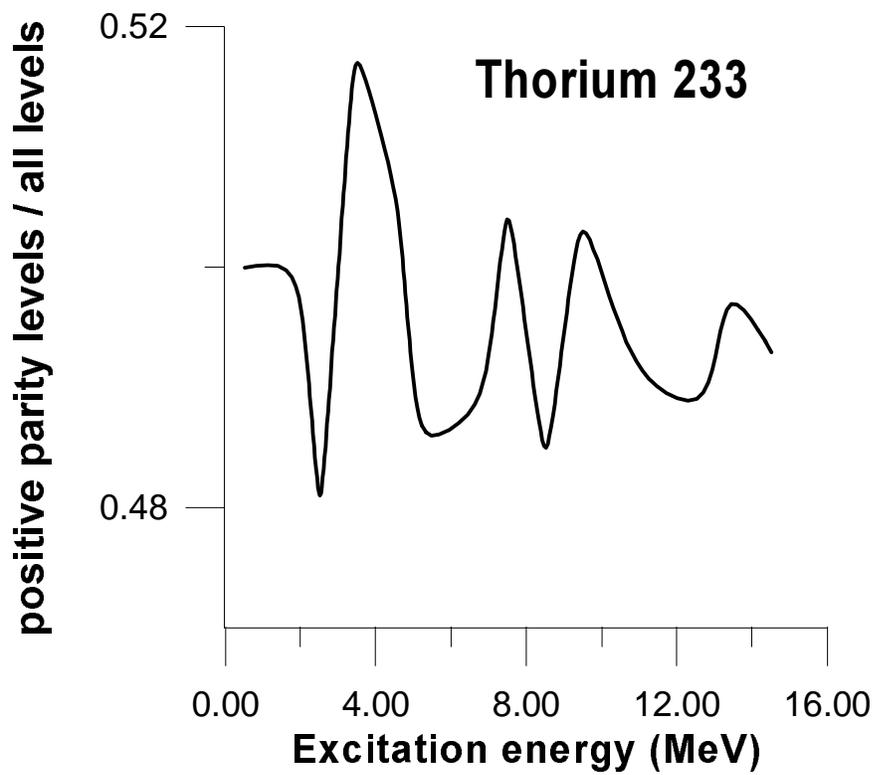
Figure 7. Percentage of positive parity levels in $^{233}$Th and $^{239}$U.
See text for details.

\newpage

\section{Tables}

\ {\bf Table 1}. 

\begin{center}
\begin{tabular}{||c|c|c|c|c|c|c||}
\hline\hline
Def. param & 1st Min. & 1 Saddle & 2d min. & 3th min. & 2 Saddle & 3 Saddle
\\ \hline
$\varepsilon $ & 0.22 & 0.383 & 0.49 & 0.7213 & 0.7044 & 0.755 \\ \hline
$\alpha _{3}$ & 0.0 & 0.0 & 0.0 & $\pm $0.1179 & 0.0 & $\pm $0.087 \\ \hline
$\alpha _{4}$ & 0.071 & -0.0827 & 0.0275 & 0.0386 & 0.0386 & 0.0386 \\ 
\hline\hline
\end{tabular}
\end{center}

\ 

\bigskip\ {\bf Table 2.} 

\begin{center}
\begin{tabular}{||c|c|c|c||}
\hline\hline
Ext. point & g$_{n}$ & g$_{z}$ & g$^{lev}$ \\ \hline
G.st & 3.651 & 2.693 & 6.345 \\ \hline
2 Min & 3.705 & 2.669 & 6.374 \\ \hline
2d Sadd Asymm & 3.77 & 2.58 & 6.35 \\ \hline
2d Sadd Symm & 4.034 & 3.069 & 7.103 \\ \hline
3rd Min & 3.661 & 2.609 & 6.27 \\ \hline\hline
\end{tabular}
\ 
\end{center}

\bigskip

\ \bigskip {\bf Table 3.} 

\begin{center}
\begin{tabular}{||c|c|c||}
\hline\hline
& Chepurnov parameters & Optimal parameters \\ \hline
Th$^{233}$ & 20.25 & 18.5 \\ \hline
U$^{239}$ & 19.9 & 18.0 \\ \hline\hline
\end{tabular}
\end{center}

\bigskip

\section{\protect\bigskip}

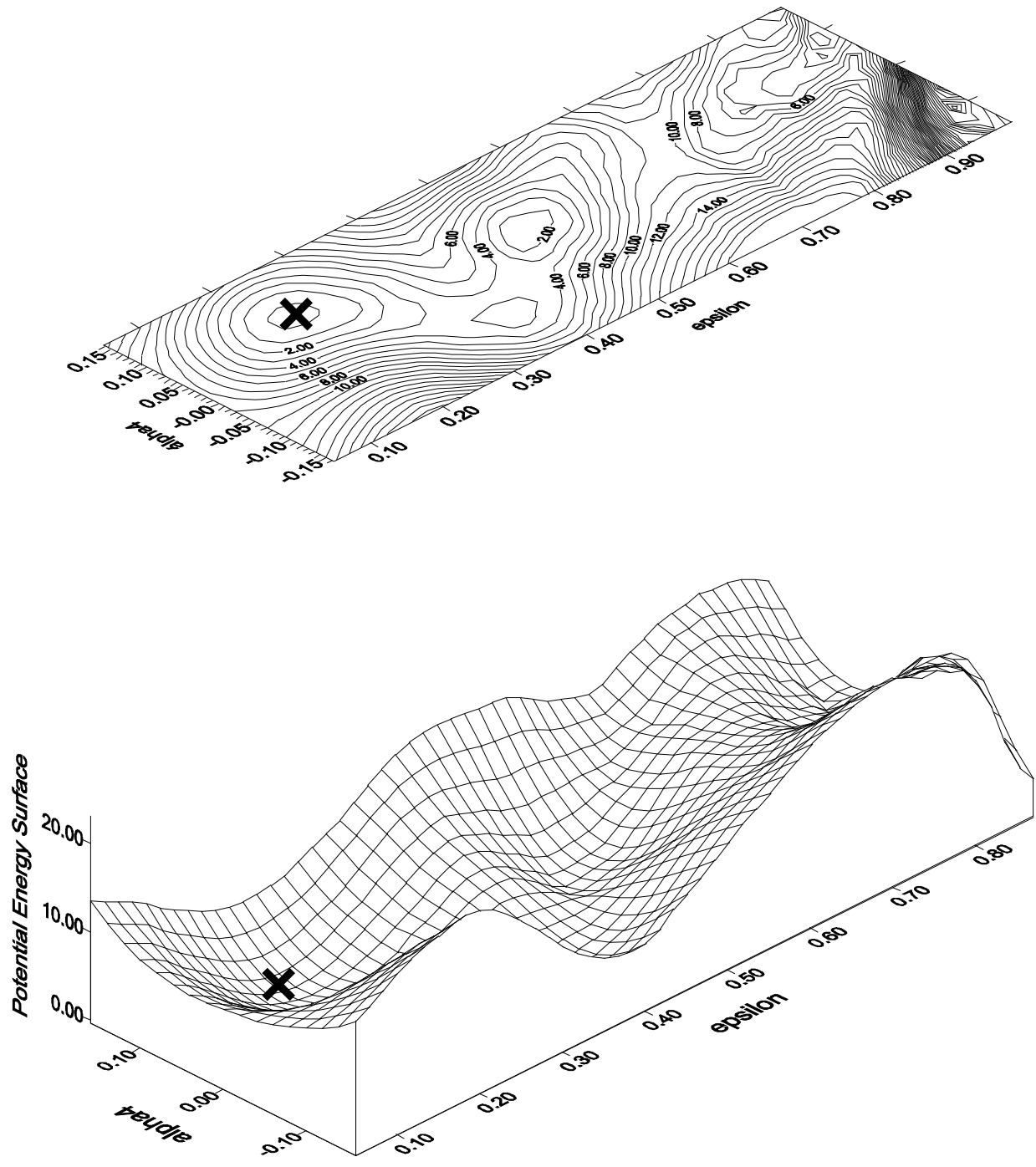
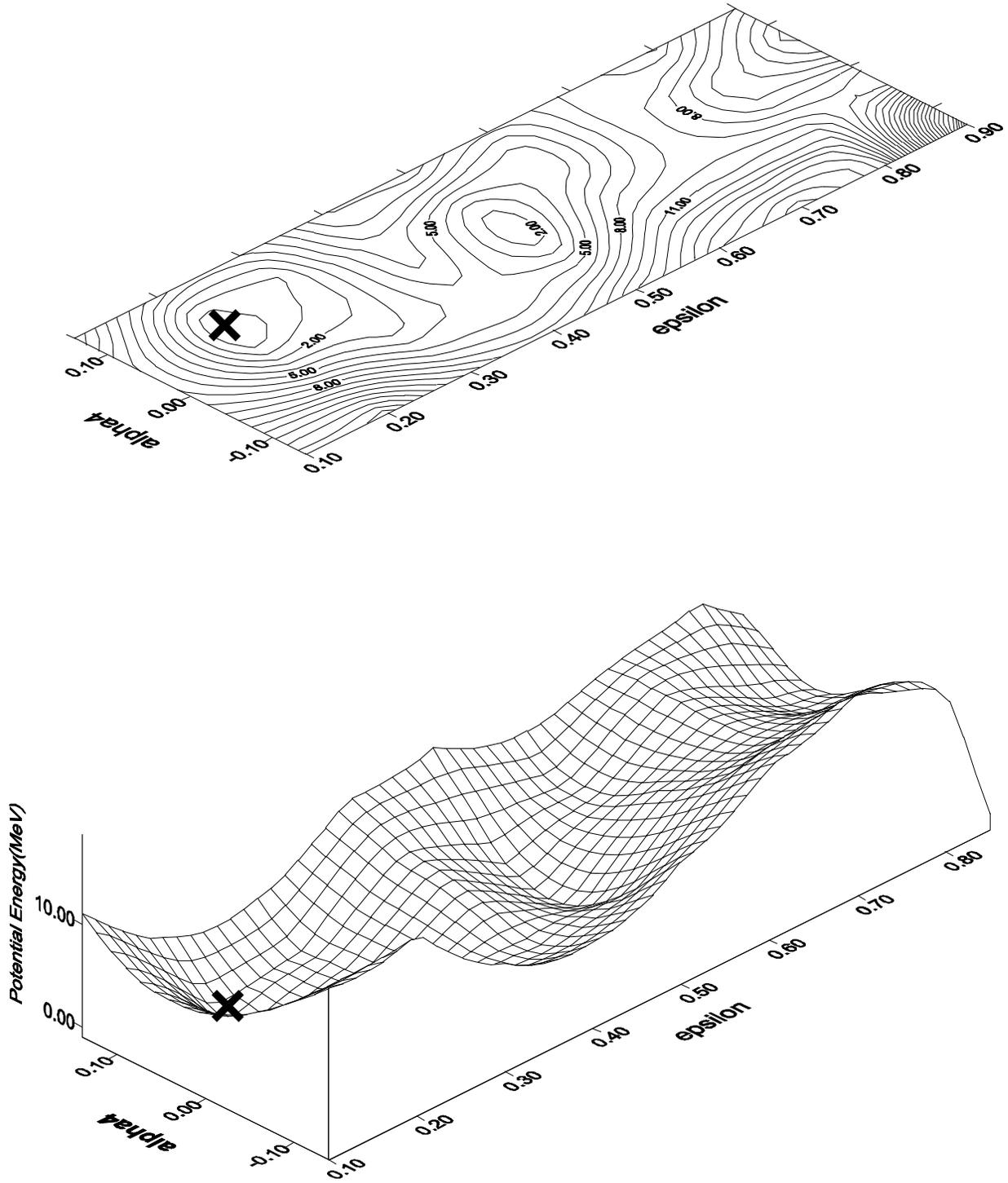
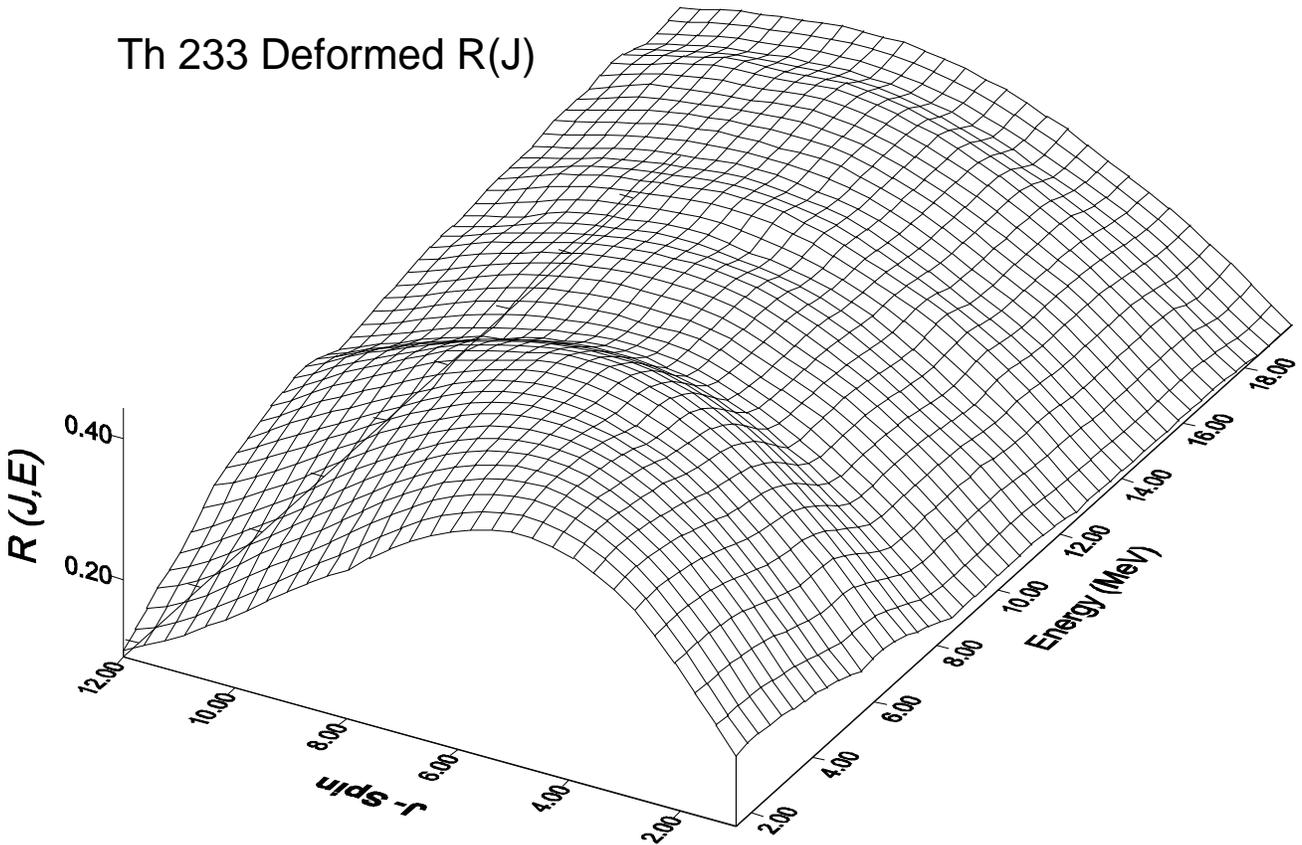
\end{document}